\documentclass[a4paper]{article}
\usepackage{fullpage}
\usepackage{graphicx}
\usepackage{amsmath,amssymb,amsthm}
\usepackage{xcolor}
\usepackage{hyperref}

\hypersetup{%
  breaklinks,%
  colorlinks=true,%
  linkcolor=[rgb]{0.45,0.0,0.0},%
  citecolor=[rgb]{0,0,0.45}
}

\newtheorem{theo}{Theorem}
\newtheorem{lem}[theo]{Lemma}
\newtheorem{corr}[theo]{Corollary}
\newtheorem{obs}[theo]{Observation}

\newtheorem*{finalthm}{Theorem~\ref{theo:final}}

\newcommand{\R}{\ensuremath{\mathbb{R}}}
\DeclareMathOperator{\INT}{int}

\newcommand{\domieq}{\succcurlyeq}

\newcommand{\D}{\mathcal{D}}

\newcommand{\cone}{\mathcal{C}}
\newcommand{\CH}{\mathcal{CH}}
\newcommand{\VC}{\mathcal{V}}
\newcommand{\eps}{\varepsilon}
\newcommand{\rank}{\eta}
\newcommand{\reg}{\mathcal{R}}

\newenvironment{denseitems}{\list{$\bullet$}%
  {\labelwidth3em\itemsep0pt\parsep0pt\topsep0.6ex}}{\endlist}

\let\geq\geqslant
\let\leq\leqslant
\let\ge\geqslant
\let\le\leqslant

\makeatletter
\partopsep\z@ \textfloatsep 10pt plus 1pt minus 4pt
\def\section{\@startsection {section}{1}{\z@}{-3.5ex plus -1ex minus
-.2ex}{2.3ex plus .2ex}{\large\bf}}
\def\subsection{\@startsection{subsection}{2}{\z@}{-3.25ex plus -1ex
minus -.2ex}{1.5ex plus .2ex}{\normalsize\bf}}
\def\@fnsymbol#1{\ensuremath{\ifcase#1\or 1\or 2\or
    3\or 4\or 5\or 6\or 7 \or 8\ or 9 \or 10\or 11 \else\@ctrerr\fi}}
\makeatother

\title{Computation of Spatial Skyline Points}

\author{Binay Bhattacharya%
  \thanks{School of Computing Science, Simon
    Fraser University, Canada. binay@cs.sfu.ca.}
  \and Arijit Bishnu%
  \thanks{Advanced Computing and
    Microelectronics Unit, Indian Statistical Institute, India.
    \{arijit, sandipdas\}@isical.ac.in.}
  \and Otfried Cheong%
  \thanks{School of Computing, KAIST,
    Korea. otfried@kaist.airpost.net.}
  \and 
  Sandip Das\footnotemark[2]
  \and 
  Arindam Karmakar%
  \thanks{Department of Computer Science and Engineering, 
    Tezpur University, India. arindam@tezu.ernet.in.}
  \and 
  Jack Snoeyink%
  \thanks{Department of
    Computer Science,  University of North Carolina at Chapel Hill,
    USA. snoeyink@cs.unc.edu.}}

\begin{document}

\maketitle
\begin{abstract}
  We discuss a method of finding skyline or non-dominated sites in a
  set $P$ of $n$ point sites with respect to a set $S$ of $m$~points.
  A site $p \in P$ is non-dominated if and only if for each $q \in P
  \setminus \{p\}$, there exists at least one point $s \in S$ that is
  closer to~$p$ than to~$q$.  We reduce this problem of determining
  non-dominated sites to the problem of finding sites that have
  non-empty cells in an additively weighted Voronoi diagram under a
  convex distance function. The weights of said Voronoi diagram are
  derived from the coordinates of the sites of~$P$, while the convex
  distance function is derived from~$S$. In the two-dimensional plane,
  this reduction gives an $O((n + m) \log (n + m))$-time algorithm to
  find the non-dominated points.
\end{abstract}

\section{Introduction}
\label{sec:intro}

Consider a hotel recommendation system for a city with many hotels,
located at point sites $P = \{p_1, \ldots, p_{n} \}$.  A tourist
proposes to visit a set of $m$~locations of interest, $S=\{
$museum~$s_1$, restaurant~$s_2$, garden~$s_3$, \dots, beach~$s_m\}$,
and would like a short list of hotels near these locations.  The
system need not list any hotel~$p \in P$ that is farther from all
locations in $S$ than some other hotel~$q \in P$.

It turns out that this is a special case of a problem considered in
the database community~\cite{bks-tso-01,ss-ssq-06}.  Consider a
database whose entries are objects with $d$~attributes of
interest. Given two objects~$p$ and~$q$, we write $p \domieq q$ if
every attribute of~$p$ is larger or equal to the corresponding
attribute of~$q$.  If $p \domieq q$ but not $q \domieq p$, then we say
that $p$ \emph{dominates}~$q$.  An object is called
\emph{non-dominated} or a \emph{skyline} object if it is not dominated
by any object in the database.  A \emph{skyline query} is the problem
of determining the skyline objects in a database with respect to a
given set of attributes.  B\"{o}rzs\"{o}nyi et al.~\cite{bks-tso-01}
proposed to add a \emph{skyline operator} to solve skyline queries in
an existing (relational, object-oriented, or object-relational)
database system.

Our hotel recommendation problem fits this framework exactly if we
choose the attributes of each point in $P$ to be the negative distances 
to the points in~$S$.  Sharifzadeh and Shahabi~\cite{ss-ssq-06} use the term
\emph{spatial skyline query} for this special version of the problem.
They suggest other application scenarios in defense or crisis
management, such as identifying a set of buildings that are to be
evacuated ahead of other buildings in case of multiple fires.

In a spatial skyline query, the distances to the points of~$S$ are
considered \emph{attributes} describing the sites of~$P$.  A
site~$p\in P$ dominates $q\in P$ if and only if it is \emph{strictly
  better} in at least \emph{one} attribute and is at least as good in
\emph{all} attributes.  In our scenario for a hotel recommendation
system, if $p \in P$ is dominated by~$q\in P$, then $p$~need not be on
the short list of hotels for a tourist visiting~$S$.  On the other
hand, if $p \in P$ is not dominated by any $q \in P$, then $p$ is a
\emph{non-dominated point site} or a \emph{skyline point}.  Examples
of skyline points are the discrete Fermat-Weber point, which is the
site in~$P$ that minimizes the sum of distances to the points~$S$, and
the sites in~$P$ that are a nearest neighbor of some point in~$S$.

\begin{figure}[ht]
  \centerline{\includegraphics[width=.4\textwidth]{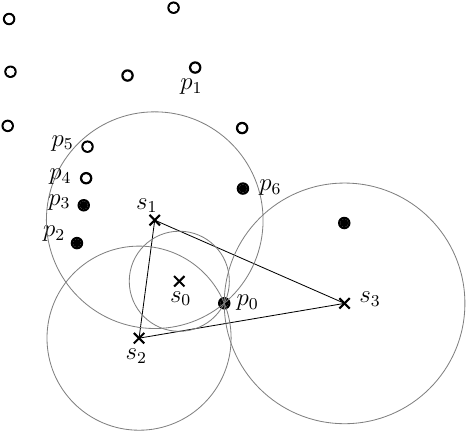}}
  \caption{Among the hotels at sites $P$ (small circles), the five
    filled circles are the \emph{skyline points} for the locations
    $S=\{s_0,\ldots,s_3\}$ (crosses). Four larger circles with centers
    in $S$, passing through $p_0$, show that most sites are dominated
    by $p_0$.  Sites $p_4$ and $p_5$ are not, but are dominated by
    $p_3$.}
  \label{fig:intro}
\end{figure}

We will use the Euclidean metric, although other metrics can be
substituted, often without changing the set of skyline points.  Since
the distance from each point $s\in S$ becomes an attribute of an
object, and objects are compared only attribute-wise, any metric that
increases monotonically with increasing Euclidean distance will
identify the same set of skyline points. In fact, after retrieving the
short list of hotels, our tourist could even use a separate
metric~$d_s$ for the distance to each site~$s\in S$ to reflect his or
her interest in each site.  Because they are defined by dominance, our
tourist is guaranteed not only that the best hotel is present on the
short list, but also that every hotel on the list is the best for some
combination of the distances.  More formally, the skyline points are
exactly the sites $p\in P$ that minimize the sum $f(p)=\sum_{s \in S}
d_s(p,s)$ for some choice of weighted distance
functions~$d_s$~\cite{bks-tso-01}. Thus, they generalize the discrete
Fermat-Weber points.

\paragraph{Prior Work.}

The skyline problem and its variants were named by researchers in the
database community~\cite{bks-tso-01,ss-ssq-06}.  B\"{o}rzs\"{o}nyi et
al.~\cite{bks-tso-01} gave details for implementing a divide and
conquer algorithm from computational
geometry~\cite{klp-fmsv-75,ps-cgi-85} in a database context to take
$O(n\log^{d-2}n +dn\log n)$ time for $d$~attributes of interest.
Since then, several other works have used nearest neighbor
search~\cite{krr-sssoasq-02}, sorting~\cite{cggl-swp-03} and index
structures~\cite{ptfs-pscds-05,teo-epsc-01}, primarily aiming to show
an experimental improvement over the results of B\"{o}rzs\"{o}nyi et
al.~\cite{bks-tso-01}.

To apply the method of B\"{o}rzs\"{o}nyi et al.~to a \emph{spatial}
skyline query with $n$~candidate sites and $m$~locations of interest
determining the attributes, one would build the attribute vectors and
then compute the skyline points in $O(n \log^{m - 2}n + mn \log n)$
time.

For the planar case, Sharifzadeh et al.~\cite{SharifzadehSK09},
correcting an error in their earlier algorithm~\cite{ss-ssq-06} that
was pointed out by Son et al.~\cite{slah-ssqaega-09}, give two
algorithms for the problem, one based on R-trees, one based on a
transversal of Voronoi diagrams and Delaunay triangulations, and
compare them experimentally.

Lee et al.~\cite{LeeSAH11} give an algorithm that runs in time $O(n (s
\log m + n))$, where $s$ is the number of reported skyline points.  In
the worst case, this is~$O(n^{2} \log m)$.  They show that both
the worst-case running time and the experimental performance of their
algorithm is significantly better than the algorithms by Sharifzadeh
et al.~\cite{SharifzadehSK09}.  They also give an approximation algorithm.

Since it appears hard to improve beyond quadratic running time under
the Euclidean metric, Son et al.~\cite{SonHA14} consider skyline
points under the Manhattan metric instead, and are able to give
an~$O((n + m)\log (n + m))$-time algorithm.

Another line of research suggests that in situations where there are
many skyline points, the user may be better served by reporting only
the top~$k$ skyline points, with respect to some scoring function.
Goncalves and Vidal~\cite{GoncalvesV09} give an algorithm that solves
this problem in~$O(n^{2} d)$ time, where $d$~is the number of
attributes defining the skyline points.  Returning to the spatial
setting, Son et al.~\cite{SonSKA17} consider the problem of reporting
the top-$k$ skyline points under the Manhattan metric, achieving
an~$O(\log n)$ improvement compared to computing all skyline points
and ranking those.

\paragraph{Results.}

We first discuss the formal definition and some degenerate situations
in Section~\ref{sec:prelim}.  In Section~\ref{sec:voronoireduce} we
then use lifting techniques~\cite{b-gtfga-80} to give several
geometric views of dominance and non-dominance problems, involving
balls, lower envelopes of cones, and Voronoi diagrams with a convex
polygonal distance function (determined by~$S$) and additive weights
(determined by~$P$).

In Section~\ref{sec:algo} we turn to the situation where the sites~$P$
and points~$S$ are given in the two-dimensional plane.  Here we can
turn our transformations into an efficient algorithm with running
time~$O((n+m)\log(n+m))$ .  This is the first algorithm to improve
upon the quadratic time barrier, and matches the bound for the
Manhattan distance by Son et al.~\cite{SonHA14}.

After the transformation, our algorithm solves the following question:
given a convex polygonal cone~$\cone$ of complexity~$m$ in~$\R^{3}$,
and $n$~translation vectors~$t_1, \dots, t_n$, decide which of the
translated cones~$\cone + t_i$ are \emph{redundant} in the sense that
they do not contribute to the union of the~$n$ cones. Since the union
of the cones is equivalent to their lower envelope, a cone~$\cone +
t_i$ is redundant exactly if its apex does not appear on the lower
envelope of the family of cones.

We make use of a recent algorithm by Biniaz et al.~\cite{biniaz-2018},
who transform a different problem (on the existence of a two-point
consistent subset) to a similar question about translates of a convex
polygonal cone.  The key idea is to represent the union of the cones
using the compact representation of Voronoi diagrams by McAllister et
al.~\cite{mks-cplvd2-96}---while the complexity of the union could
be~$\Theta(nm)$, the compact representation has
complexity~$O(n)$---and to use a simplified version of their
sweep-line algorithm for the computation of Voronoi diagrams to answer
our question.\footnote{In the conference version~\cite{walcom} of this
  paper, we originally suggested using the randomized incremental
  algorithm of McAllister et al.~\cite{mks-cplvd2-96}.  It appears,
  however, that this does not lead to an algorithm with the claimed
  running time, as the history-DAG built by the algorithm does not
  necessarily have constant out-degree. The interested reader may want
  to look at Fig.~14(a) in~\cite{mks-cplvd2-96}: in the depicted
  situation, it is true that the spoke region~$A$ intersects only two
  spoke regions in the updated diagram.  However, when~$P$ lies
  inside~$A$ but conflicts with the two vertices~$u$ and~$v$, then
  every spoke region incident to~$P$ will intersect~$A$, and the
  number of such regions is only bounded by the complexity of the cell
  of~$P$.}

\section{Preliminaries}
\label{sec:prelim}

Throughout the paper, $P$~will denote a set of $n$ point sites,
and~$S$ will denote a set of $m$~points (locations) in~$\R^d$.  We
use~$d(x,y)$ for the usual $L_2$-distance in~$\R^d$.  For a set $X
\subset \R^d$, we denote its \emph{interior} as~$\INT X$.

For two sites $p, q \in P$ we write $p \domieq q$ if $d(p, s) \leq
d(q, s)$ for every~$s \in S$.  A site $p \in P$ dominates~$q\in P$ if
$p \domieq q$ but not $q \domieq p$, which is equivalent to saying
that we have $d(p, s) \leq d(q,s)$ for \emph{all} $s \in S$ and
$d(p,s) < d(q, s)$ for \emph{at least one}~$s \in S$.  If $p \in P$ is
not dominated by any $q \in P$, then $p$ is a \emph{non-dominated
  point site} or a \emph{skyline point}.

Let us first investigate the case of two sites $p, q \in P$ with $p
\domieq q$ and $q \domieq p$. This implies that $d(p,s) = d(q,s)$ for
all points in~$S$. In other words, all points in~$S$ lie on the
bisector of~$p$ and~$q$, and therefore in a flat of dimension at
most~$d-1$.  We can ensure that this case does not arise by a simple
preprocessing of the point set: Let~$H$ be the affine hull of~$S$.  If
$H$ has dimension~$k < d$, then we replace $P$ by a point set in $k+1$
dimensions that preserves all distances to the flat~$H$ and such that
$P$ is contained in one closed halfspace of~$\R^{k+1}$ bounded by~$H$.
With this preprocessing, we will have that for any two sites $p \neq
q$ in~$P$, we always have $d(p,s )\neq d(q,s)$ for some~$s \in S$, and
we will make this assumption throughout the paper. Note that the
preprocessing may map several sites of~$P$ into the same
point---either all the original sites are skyline points, or none of
them is.

With this assumption, $p \in P$ dominates $q \in P \setminus\{p\}$
simply if $p \domieq q$, that is if $d(p, s) \leq d(q, s)$ for all $s
\in S$.  Conversely, $p$ is \emph{not dominated} by~$q$ if and only if
there is a site~$s \in S$ such that $d(p, s) < d(q, s)$.  It follows
that a site $p \in P$ is a skyline point if and only if for every site
$q \in P \setminus \{p\}$ there is a site $s_q \in S$ with $d(p, s_q)
< d(q, s_q)$.

Our problem is to extract the skyline points of~$P$ with respect
to~$S$.  Let $h(p, q)$ denote the half-plane containing~$p$ that is
bounded by the bisector of~$p$ and~$q$.  A brute force approach to
identify whether $p \in P$ is a skyline point is to determine, for all
$q \ne p$, if at least one site $s \in S$ lies in $h(p, q)$.  This
takes $\Theta(m n)$ time for each~$p$, giving a total time
of~$\Theta(mn^2)$.

\section{Views of non-dominance problems in~$\R^d$}
\label{sec:voronoireduce}

We first relate point domination in $d$~dimensions to balls (disks as
in Figure~\ref{fig:intro}), then to the envelope of cones, and finally
to the additively weighted Voronoi diagram of a convex distance
function.  We will define these terms as we go, culminating in the
following Theorem~\ref{theo:final}, which we prove at the end of the
section.  
\begin{finalthm}
  The skyline or non-dominated points of a set~$P \subset \R^d$ with
  respect to locations~$S \subset \R^d$ are those with non-empty
  Voronoi cells under a convex distance function determined by~$S$
  with additive weights determined by~$P$.
\end{finalthm}
In Section \ref{sec:algo}, we will show that for sites and locations
\emph{in the plane}, the reduction implied by Theorem~\ref{theo:final}
leads to an efficient algorithm.

\subsection{Dominated points and balls}
\label{ssec:diskdomination}

\begin{figure}[ht]
  \centerline{\includegraphics[width=.4\textwidth,page=2]{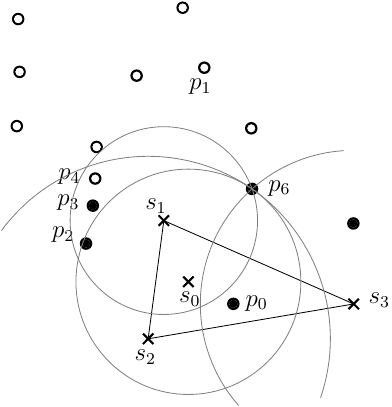}}
  \caption{Site~$p_6$ is not dominated.}
  \label{fig:not-dominated}
\end{figure}

For $x, y \in \R^{d}$, let $C(x,y)$ denote the closed ball with
center~$x$ and radius~$d(x,y)$.  For a site~$p \in P$, consider the
balls~$C(s, p)$ centered at each~$s \in S$.  Figure~\ref{fig:intro}
illustrates that the sites dominated by $p_0$ are outside the union
of the disks through~$p_0$, that is, outside $\bigcup_{s \in S} C(s,
p_0)$.  On the other hand, Figure~\ref{fig:not-dominated} illustrates
that $p_6$ is not dominated, because no site of $P$ is inside the
\emph{intersection} of its disks.  These two complementary views of
dominance and non-dominance can be contrasted throughout the entire
Section~\ref{sec:voronoireduce}.

For $p \in P$, we define the \emph{dominator region} of $p$ as $D_p =
\bigcap_{s \in S} C(s, p)$, and the \emph{dominated region} of $p$ as 
$\D_p = \R^{d} \setminus \bigcup_{s \in S} \INT C(s, p)
= \bigcap_{s \in S} \big(\R^{d} \setminus \INT C(s, p)\big)$.  
We make the following observations:
\begin{obs}
  \label{obs1}
  For site $p\in P$ we have:
  \begin{description}
  \item[\textmd{(i)}] 
    a site $q \in P$ dominates $p$ if and only if $q \in D_p$;
  \item[\textmd{(ii)}] 
    $p$ dominates a site $q \in P$ if and only if $q \in \D_p$;
  \item[\textmd{(iii)}]
    $p$ is a skyline point if and only if $D_p$ does
    not contain any site $q \in P \setminus\{p\}$;
  \item[\textmd{(iv)}]
    $p$ is a skyline point if and only if $p \not\in \D_q$ for all $q
    \in P\setminus\{p\}$;
  \item[\textmd{(v)}] if a site $q \in P$ lies in $D_p$ then $D_q
    \subset D_p$;
  \item[\textmd{(vi)}] if a site $q \in P$ lies in $\D_p$, then $\D_q
    \subset \D_p$;
  \item[\textmd{(vii)}] $D_p$ is a non-empty convex region bounded by
    spherical patches centered at vertices of the convex hull of
    sites, ${\cal CH}(S)$.
  \end{description}
\end{obs}

\begin{proof}
  We have $q \domieq p$ if and only if $d(q,s)\leq d(p,s)$ for all $s
  \in S$, which is equivalent to $q \in C(s,p)$ for all~$s \in S$, or
  $q \in D_p$, implying claim~(i).

  On the other hand $p \domieq q$ if $d(p,s) \leq d(q,s)$ for all~$s
  \in S$. This is equivalent to $q \not\in \INT C(s, p)$ for all $s
  \in S$, or $q \not \in \bigcup_{s\in S}\INT C(s, p)$, implying
  claim~(ii).

  Claims~(iii) and~(iv) now follow immediately from the definition of
  skyline points.

  For (v), $q \in D_p$ implies $q \domieq p$, or $d(q, s)\leq d(p, s)$
  for all $s \in S$.  Then $C(s,q)\subseteq C(s,p)$ for all $s\in S$,
  and so $D_q \subset D_p$.  

  For (vi), $q \in \D_p$ implies $p \domieq q$, or $d(p,s)\leq d(q,s)$
  for all $s \in S$.  Then $\INT C(s, p) \subseteq \INT C(s, q)$, and
  $\bigcup_{s\in S}\INT C(s,p)\subset \bigcup_{s\in S}\INT C(s,q)$.
  This implies $\D_p \supset \D_q$.

  For~(vii) we observe that $D_p$ is the intersection of balls and therefore
  convex and bounded by spherical patches; since all balls contain~$p$, the
  intersection is not empty. Fix a unit vector~$v$ and consider where the
  ray from $p$ in direction~$v$ leaves~$D_p$:
  $\max_{\alpha}(p+\alpha v) \in D_p$.  
  For each inequality $d(p+\alpha v,s) \le  d(p,s)$, we can
  square both sides and rewrite as
                       $\alpha \le 2(s-p)\cdot v$.  
  Thus, $\alpha$ is determined by the extreme site~$s$ in direction~$v$, 
  and this site is on the convex hull,~$\mathcal{CH}(S)$.
\end{proof}
In the plane, the boundary of the dominator region~$D_p$ is determined
by at most~$m$ circular arcs, and so the total complexity of the
dominator regions for all the sites in~$P$ is~$\Theta(m n)$.

Since the dominated region is non-convex, it looks more complex to
work with, but in fact this distinction will disappear as we lift to
cones in the next subsection.

\subsection{Dominator cones and dominated cones}
\label{ssec:conedominate}

We use Brown's lifting map to generate cones from
balls~\cite{b-gtfga-80}.  Assume a coordinate system with origin
inside the convex hull~$\CH(S)$. Consider sites and locations in
the plane for the moment, and lift $S$ and $P$ to the unit paraboloid
$\Psi = \{(x,y,z) \mid z = x^2 + y^2 \}$; a point $p=(x,y)$ in the
plane is lifted to the point $p'=(x, y, x^2+y^2)$ on~$\Psi$.  Note
that lifting a circle $C=\{(x-c_1)^2+(y-c_2)^2=r^2\}$ gives a set
$C'=\{z-2c_1x-2c_2y+c_1^2+c_2^2-r^2=0\}$ on~$\Psi$ that is linear in
$x$, $y$, and $z$~\cite{su-hocg-00}. Thus, we can consider $C'$ to be
a plane in $3$-dimensional space.  Points inside the circle~$C$ are
lifted to points on the paraboloid~$\Psi$ that lie in the halfspace
below~$C'$, which we denote~$C^-$.  Points outside the circle map to
points on~$\Psi$ that lie in the halfspace above, denoted~$C^+$.

Lifting in higher dimensions is analogous, with $p'$ adding a final
dimension of $p\cdot p$ to a point~$p \in \R^{d}$, and spheres being
lifted to hyperplanes.  For a ball~$C(s, p)$, we denote the lifted
hyperplane as~$C'(s,p)$, the \emph{closed} halfspace below this
hyperplane as~$C^{-}(s,p)$, and the closed halfspace above the
hyperplane as~$C^{+}(s, p)$.  

For $p \in P$, we define the \emph{dominator cone} $\Lambda_p =
\bigcap_{s \in S} C^{-}(s, p)$, and the \emph{dominated cone} $V_{p} =
\bigcap_{s \in S} C^{+}(s, p)$.  They are directly related to the
dominator region and dominated region as follows:
\begin{obs}
  \label{obs2}
  For points $q \in \R^{d}$ and $p \in P$, we have $q \in D_{p}$ if and
  only if $q' \in \Lambda_{p}$, and $q \in \D_{p}$ if and only if $q'
  \in V_{p}$.
\end{obs}
\begin{proof} 
  We have $q \in C(s,p)$ if and only if $q' \in C^{-}(s, p)$.  It
  follows that $q \in D_{p} = \bigcap_{s \in S}C(s, p)$ if and only if
  $q' \in \bigcap_{s \in S} \Lambda_{p} = C^{-}(s, p)$.  Similarly $q \not\in
  \INT C(s, p)$ if and only if $q' \in C^{+}(s, p)$.  It follows that
  $q \in \D_{p} = \bigcap_{s \in S} \big(\R^{d} \setminus \INT
  C(s,p)\big)$ if and only if $q' \in V_{p} = \bigcap_{s \in S}
  C^{+}(s, p)$.
\end{proof}

Let $o$ denote the origin of the coordinate system, so that $o' = o$,
and consider the cones $\Lambda = \bigcap_{s \in S} C^{-}(s, o)$ and
$V = \bigcap_{s \in S} C^{+}(s, o)$.  Since $C'(s, o)$ passes
through~$o$, both cones have their apex in~$o$, and since $C^{+}(s, o)
= -C^{-}(s, o)$, we have $V = -\Lambda$.

Now we observe that the hyperplane $C'(s, p)$ is parallel to the
hyperplane tangent to the unit paraboloid in~$s'$, and passes
through~$p'$.  It follows that $C'(s, p) = C'(s, o) + p'$, and so
$\Lambda_{p} = \Lambda + p'$ and $V_{p} = V + p'$.  In particular, all
dominator cones are translates of~$\Lambda$---and all dominated cones
are translates of~$V$.  Both $\Lambda_{p}$ and $V_{p}$ have their apex
in~$p'$.

Consider two points $p, q \in P$.  By Observations~\ref{obs1}(i)
and~\ref{obs2}, $q$ dominates~$p$ if and only if $q' \in \Lambda_p$.
Since $\Lambda_{p}$ and $\Lambda_{q}$ are homothets, this is
equivalent to $\Lambda_{q} \subset \Lambda_p$. By the same reasoning,
$q$ dominates~$p$ if and only if $V_p \subset V_q$.  (Note that these
properties imply Observations~\ref{obs1}(v)--(vi).)

The union of a set of upward-pointing cones is known as its
\emph{lower envelope}; Figure~\ref{fig:genVor}(right) shows an
example.  We say that cone~$V_p$ is \emph{redundant} if the union of
the cones~$V_q$, for~$q \in P \setminus\{p\}$, already contains~$V_p$.
\begin{corr}
  The skyline points among the sites~$P$ with respect to the
  points~$S$ correspond exactly to the non-redundant cones in the
  lower envelope.
  \label{cor:skyline}
\end{corr}
\begin{proof}
  Since all cones are homothets, a cone $V_p$ is redundant if and only
  if $p' \in V_q$, for some $q \in P \setminus \{p\}$. We have $p' \in
  V_q$ if and only if $q$ dominates~$p$, so the claim follows.
\end{proof}

\subsection{Additively weighted Voronoi diagrams under
  a convex distance function}
\label{ssec:relation}

Constructing the lower envelope of the cones~$V_p$ is costly; the best
algorithm known would take $\Theta(nm\log (m+n))$ time even in
dimension~two.  However, lower envelopes of cones can be interpreted
as Voronoi-diagrams, and this will lead us to a more efficient method
of finding the non-redundant cones.

Minkowski showed that any compact convex set~$M$ whose interior
contains the origin defines a \emph{convex distance
  function}~$d_M(p,q)$, where the distance from point~$p$ to~$q$ with
respect to~$M$ is the amount that~$M$ must be scaled to
include~$q-p$. Mathematically,
\[
d_M(p,q) = \min \{\lambda\ge 0 \mid q-p\in\lambda M\}
\]
A convex distance function may not be a metric, since $d_M$ is
symmetric only if~$M$ is centrally symmetric: we have $d_M(p,q) =
d_{-M}(q,p)$. However, the distance function~$d_M$ satisfies the
triangle inequality~\cite{schneider2014convex}: $d_M(p,q)+d_M(q,r) \ge
d_M(p,r).$ The boundary of~$M$ serves as the unit ball for the
distance function~$d_M$.  For a fixed~$p$, the graph of $x \mapsto
d_M(p, x)$ is a cone with apex at~$p$, and every horizontal
cross-section is a homothet of~$M$. Note that the Euclidean metric is
the convex distance function with~$M$ the unit-radius ball.

Given a finite set of points $P \subset \R^d$ with additive weights
$\omega_p$ for each $p \in P$, and a convex distance
function~$d_M(p,q)$, we define the \emph{Voronoi cell} of a point~$p
\in P$ as
\[
\VC(p) = \Big\{ x \in \R^d \Bigm\vert d_M(p,x) + \omega_P < d_M(q, x) +
\omega_q \text{~for all~} q \in P \setminus \{p\} \Big\}.
\]
The \emph{Voronoi diagram} of~$P$ is the family of all Voronoi
cells~$V(p)$, for~$p \in P$.\footnote{Note that when the unit ball~$M$
  is polygonal, the Voronoi diagram may be degenerate.  The classic
  example considers the $L_1$-metric (here, $M$ is the convex hull of
  the points~$(0, 1)$, $(1,0)$, $(-1,0)$, and $(0, -1)$), and point
  sites on the line~$y = x$.  Many points have multiple equidistant
  nearest sites, and therefore do not lie in any Voronoi cell.  As a
  result, the closures of the Voronoi cells do not cover the plane.}

\begin{figure}[ht]
  \centerline{\includegraphics[width=.48\textwidth]{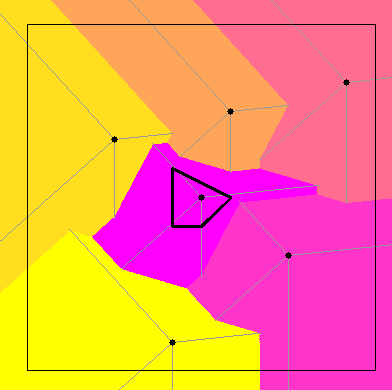}\hfill
    \includegraphics[width=.48\textwidth]{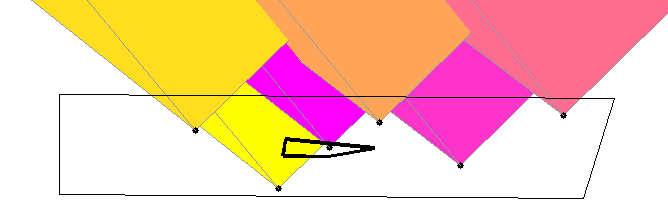}}
  \caption{A Voronoi diagram of 6 sites in the plane using a convex
    quadrilateral as distance function; its view as a lower envelope of
    cones. All weights are zero to make the 3d figure easier to
    interpret.}
  \label{fig:genVor}
\end{figure}

Figure~\ref{fig:genVor}(left) shows the Voronoi diagram of six
distinct sites in the plane, all having weight zero.  The distance
function is defined by the black convex quadrilateral around the point
at the origin, and each Voronoi cell is drawn in a different shade.

Every Voronoi diagram is the \emph{minimization diagram} of a family
of functions.  We define~$f_p(x) = d_M(p, x) + \omega_p$ for~$p \in
P$, and consider the lower envelope of these functions, that is, the
graph of the function~$f_{\min}: x \mapsto \min_{p \in P} f_{p}(x)$.
The Voronoi cell of~$p$ is the set of all those~$x$ where $f_p(x) =
f_{\min}(x)$ and $f_{q}(x) > f_{\min}(x)$ for $q \neq p$.  In this
sense, the Voronoi diagram corresponds to the projection of the graph
of~$f_{\min}$. Since the graph of each~$f_{p}$ is a cone with apex
at~$(p, \omega_p)$, the Voronoi diagram corresponds to the lower
envelope of these cones, see Figure~\ref{fig:genVor}(right) for an
illustration.
\begin{theo}
  The skyline or non-dominated points among the sites~$P$ with respect
  to points~$S$ are those with non-empty Voronoi cells under a convex
  distance function determined by~$S$ with additive weights determined
  by~$P$.
  \label{theo:final}
\end{theo}
\begin{proof}
  Consider the $(d+1)$-dimensional cone~$V = \bigcap_{s \in S}
  C^{+}(s, o)$.  The intersection of this cone with the plane $x_{d+1}
  = 1$ is a $d$-dimensional convex polytope~$M$ that contains the
  point $(0,0, \dots, 0, 1)$.  The boundary of~$V$ is therefore the
  graph of the function~$x \mapsto d_{M}(o, x)$.  Since for $p \in P$,
  we have $V_{p} = V + p'$, the boundary of $V_{p}$ is the graph of $x
  \mapsto f_p(x) = d_{M}(p, x) + \omega_p$, where $\omega_p = p \cdot
  p$.

  It follows that the lower envelope of the cones~$V_{p}$ corresponds
  to the Voronoi diagram of~$P$ with additive weights~$\omega_{p} = p
  \cdot p$ and convex distance function~$d_{M}$, where~$M$ is defined
  entirely by the points~$S$.

  A point~$p \in P$ is a skyline point if~$V_{p}$ is not redundant,
  which is equivalent to~$p'$ not lying in any cone~$V_{q}$, for $q
  \in P \setminus \{p\}$.  If this is the case, then~$f_{p}(p) <
  f_q(p)$ for $q \in P \setminus \{p\}$, and so the Voronoi cell
  of~$p$ contains at least~$p$ itself and is not empty.  If~$p' \in
  V_q$ for some~$q \neq p$, then we have $V_{p}\subset V_q$. This
  implies~$f_{p}(x) \geq f_{q}(x)$ for all~$x$, and so the Voronoi
  cell of~$p$ is empty.
\end{proof}

\section{Computing the non-redundant cones}
\label{sec:algo}

We now focus on sites and locations in the plane, because in this
important case the reductions of the previous section can be turned
into an efficient algorithm.

By Theorem~\ref{theo:final}, it would suffice to compute a Voronoi
diagram and check which sites define non-empty Voronoi cells.
Unfortunately, to our knowledge, Voronoi diagrams under convex
distance functions with additive weights have not been studied in the
literature~\cite{f-vddt-04}.

Fortunately, Biniaz et al.~\cite{biniaz-2018} recently showed how to
solve a specific problem about such diagrams: Given $n$~sites with
additive weights, a convex distance function defined by a
convex~$n$-gon, and $n$~query points, they determine the Voronoi cell
containing each query point.  They use a simplified version of the
sweepline algorithm by McAllister et al.~\cite{mks-cplvd2-96} to solve
this problem without actually computing the Voronoi diagram.  As they
already point out, their algorithm can easily be adapted to solve our
problem of determining the non-empty Voronoi cells.  In the following
we sketch their algorithm as applied to our problem, referring the
reader to Biniaz et al.~\cite{biniaz-2018} and McAllister et
al.~\cite{mks-cplvd2-96} for details.

We start by picking a coordinate system such that no two sites have
the same $x$-coordinate.  We then process the points~$S$ in~$O(m \log
m)$ time to compute the cone~$V = \bigcap_{s \in S} C^{+}(s, o)$.  A
horizontal cross section of~$V$ is a convex polygon whose edges
correspond to the extreme points in~$S$, so it has at most~$m$ edges.

By Theorem~\ref{theo:final}, our goal is to determine which elements
of the family of cones~$\{ V + p' \mid p \in P\}$ are non-redundant in
the lower envelope of the family.  We split the cone $V$ along the
$yz$-plane, resulting in a cone $V'$ in the half-space $x \geq 0$, and
a cone $V''$ in the half-space~$x \leq 0$.  We define $V_p = V + p'$,
$V'_p = V' + p'$, and $V''_p = V'' + p'$.  Since $V_{p} \subset V_q$
is equivalent to $p' \in V_{q} = V'_{q} \cup V''_q $, $V_{p}$ is
redundant if and only if $V'_{p}$ is redundant in the family~$\{ V'_p
\mid p \in P\}$ or $V''_{p}$ is redundant in the family~$\{ V''_p \mid
p \in P\}$.

We explain how to determine the non-redundant cones in the lower
envelope of the family~$\{V'_p \mid p\in P\}$.  The treatment of the
family~$\{V''_p\mid p \in P\}$ is symmetrical.  To make the lower
envelope well-behaved, we remove all degeneracies by perturbing the
additive weights~$\omega_p$ slightly.  To this end, we define the
\emph{rank}~$\rank(p)$ of~$p \in P$ as the number of points~$q \in P$
with $q_{x} \leq p_{x}$.  We have~$1 \leq \rank(p) \leq n$, and since
no two points have the same $x$-coordinate, all ranks are distinct.
We perturb $\omega_{p}$ to $\omega_{p} - \eps^{\rank(p)}$, for an
infinitesimal~$\eps > 0$.  With this perturbation, we achieve that
(i)~no apex of a cone lies on the boundary of another cone, (ii)~any
three cone boundaries have a finite number of points in common, and
(iii)~no four cone boundaries have a point in common.  Consider now a
pair of cones with $V'_{p} \subset V'_{q}$. This is equivalent to $p'
\in V'_{q}$, which implies that~$\rank(p) > \rank(q)$.  We therefore
have $\eps^{\rank(p)} \ll \eps^{\rank(q)}$, and so the perturbation
preserves the containment.  It follows that the set of non-redundant
cones remains unchanged by the perturbation.

As before, we interpret the boundary of the cone~$V'_p$ as the graph
of a function~$f_{p}$.  The function~$f_{p}$ is now a partial
function, defined only on the half-plane~$x \geq p_x$.  We define
$f_{\min}(x, y) = \min_{p\in P, p_x \leq x} f_{p}(x, y)$, and the
regions~$\reg_{p}$ where~$f_{p}$ gives the minimum:
\[
\reg_{p} = \bigl\{ (x, y) \bigm\vert f_{p}(x, y) = f_{\min}(x, y) \bigr\}.
\]
Because of our perturbation, the regions~$\reg_{p}$ are
interior-disjoint, and their closures cover the half-plane bounded by
the vertical line through the leftmost site. A cone $V'_{p}$ is
redundant if and only if~$\reg_{p}$ is empty.  Biniaz et
al.~\cite{biniaz-2018} observe the following folklore properties of the
regions~$\reg_{p}$:
\begin{denseitems}
\item $\reg_{p}$ is star-shaped with respect to~$p$: For every~$x \in
  \reg_{p}$, the segment~$px$ lies in~$\reg_{p}$.
\item For three distinct sites~$p, q, r \in P$, the
  intersection~$\reg_p \cap \reg_q \cap \reg_r$ contains at most two
  points.
\end{denseitems}
We add the following observation:
\begin{lem}
  \label{lem:x-monotone}
  For two distinct sites~$p, q \in P$ the intersection~$\reg_{p} \cap
  \reg_{q}$ is $x$-monotone in each half-plane bounded by the line~$pq$.
\end{lem}
\begin{proof}
  We can assume $\rank(p) < \rank(q)$ and consider the half-plane
  above the line~$pq$.  Assume there are two points~$s_1 = (x, y_1)$
  and~$s_2 = (x, y_2)$, with $y_1 < y_2$, that lie in~$\reg_{p} \cap
  \reg_{q}$.  Then the segment~$ps_1$ lying in~$\reg_p$ and the
  segment~$qs_2$ lying in~$\reg_q$ intersect, a contradiction.
\end{proof}
Biniaz et al.~\cite{biniaz-2018} show how to
adapt~\cite[Lemma~3.15]{mks-cplvd2-96}, which makes use of Kirkpatrick
and Snoeyink's tentative prune-and-search
technique~\cite{ks-tpscf-95}, to perform each of the following
computations in~$O(\log m)$ time:
\begin{denseitems}
\item Given a point~$(x,y)$ and~$p \in P$, evaluate~$f_{p}(x, y)$;
\item Given three distinct sites~$p, q, r \in P$, compute~$\reg_{p}
  \cap \reg_q \cap \reg_r$;
\item Given two distinct sites~$p, q \in P$ and a vertical
  line~$\ell$, compute~$\ell \cap \reg_{p} \cap \reg_{q}$.
\end{denseitems}
Armed with these primitives, we can now explain the algorithm.  We
sweep a vertical line~$\ell \equiv \bigl\{ (x, y) \bigm\vert x =
t\bigr\}$, where $t$ goes from~$-\infty$ to~$+\infty$.  During the
sweep, we maintain the sequence of regions~$\reg_p$ intersected by the
sweepline, in order of increasing $y$-coordinate.  A fixed
region~$\reg_p$ may appear several times in this sequence.  For each
element of the sequence except the topmost and bottommost one, we
maintain a \emph{spoke}: a non-vertical line segment or half-line
intersecting the sweepline inside the region.  More precisely,
when~$\reg_p$ appears on the sweepline with upper neighbor~$\reg_q$
and lower neighbor~$\reg_r$, then the spoke connects~$p$ with the
``vertex'' $v = \reg_q \cap \reg_p \cap \reg_r$ where $q, p, r$ appear
counter-clockwise in this order around~$v$.  If no such point~$v$
exists, then the spoke extends up to infinity.

When there are~$k$ elements in the sweepline sequence, we thus
have~$k-2$ spokes that separate the sweepline into~$k-1$ intervals.
Each interval belongs to two adjacent regions.

Since the bisectors~$\reg_p \cap \reg_q$ are $x$-monotone, the
sequence of regions intersecting the sweepline changes in only two
ways: At a \emph{site event}, the sweep line reaches one of the
sites~$P$, which can cause a new region to appear.  At a \emph{vertex
  event}, we reach (locally) the end of a region. The point where this
happens is necessarily the endpoint of the spoke for the region.

The sweep is initialized by adding the first three sites to the
sweepline status. In $O(\log m)$ time we determine the order of the
regions along the sweepline and compute the spoke for the middle
element.  We add the spoke endpoint as a vertex event to the event
queue.

At a \emph{vertex event}, we first check if the regions~$\reg_p,
\reg_q, \reg_r$ defining the spoke endpoint are still adjacent on the
sweepline.  If not, we can ignore the event.  Otherwise, we
remove~$\reg_q$ from the sweepline status.  This causes the lower
neighbor of~$\reg_p$ and the upper neighbor of~$\reg_r$ to change, so
we recompute their spokes and add the new spoke endpoints to the event
queue.  All this takes $O(\log n + \log m)$ time.

Consider now a \emph{site event}, where the sweepline reaches site~$p
\in P$.  We perform binary search on the spokes to locate the interval
between two spoke segments containing~$p$.  Since the spoke segments
are given geometrically, this takes~$O(\log n)$ time.  Once we know
that $p$ lies in one of the two regions~$\reg_q$ or~$\reg_r$, we
evaluate~$f_q(p)$ and~$f_{r}(p)$. Let's say that $f_{q}(p) <
f_{r}(p)$, so that $p$ lies in~$\reg_q$.  If $f_{q}(p) < \omega_p$,
then~$\reg_p$ is empty and $V'_p$ is redundant, and we are done with
the site.  Otherwise, $V'_p$ is not redundant, and $\reg_p$ needs to
be inserted into the sweepline status.  We duplicate the region
for~$\reg_q$ and insert~$\reg_p$ between the two copies.  We then walk
up and down the sweepline status from~$p$ to determine whether and
which regions disappear from the sweepline status.  We can decide that
for a region~$\reg_s$ by computing its intersection with the sweepline
in~$O(\log m)$ time, and evaluating~$f_s$ and~$f_{p}$ at those points.
Finally, we compute the spoke segment for~$\reg_p$ and recompute it
for the two neighbors of~$\reg_p$ in the sweepline status, inserting
the spoke endpoints into the event queue.  Handling the event thus
takes time~$O(\log n + (k+1)\log m)$, where~$k$ is the number of
regions removed from the sweepline status.

A site event increases the size of the sweepline status by at most
two, and creates at most three vertex events.  A vertex event creates
up to two new vertex events, but only if it removes a region from the
sweepline status.  We can thus charge the vertex events to the at
most~$2n$ regions ever appearing on the sweepline status, and observe
that both sweepline status and event queue have size~$O(n)$.  The
total running time is thus $O\bigl(m \log m + n (\log n + \log m)\bigr) =
O\bigl((n+m) \log (n+m)\bigr)$.

We summarize this section with the following theorem.
\begin{theo}
  The skyline or non-dominated points among $n$~sites in the plane
  with respect to a set of $m$~locations can be computed in
  time~$O\bigl((n+m) \log (n+m)\bigr)$.
\end{theo}

\section{Conclusions}
\label{sec:conclude}

In this paper, we proposed an algorithm for finding the non-dominated
or skyline points among a point set~$P$ with respect to a set of
sites~$S$ in $\R^2$. This problem was initially proposed by
Sharifzadeh and Shahabi~\cite{ss-ssq-06} and termed the
\emph{spatial skyline query} problem. We gave some geometric
insights into this problem to design an efficient algorithm to find
the skyline points, especially when the points and sites are given in
the plane.

Since our reductions apply in all dimensions, it would be interesting
to extend the algorithms to dimensions higher than two, although for
geographical applications, the 2-d map is the most relevant. It may be
more important to consider dynamic versions of spatial skyline queries
that support insertion and deletion of sites and data points, and to
combine 2-d spatial skyline queries with one or more non-spatial
attributes.  The decomposability of skyline queries means that there
are interesting trade-offs to be considered when partitioning the
problem on spatial or non-spatial dimensions.

\end{document}